\documentstyle[12pt]{article}
\begin{document}
\begin{flushright}
{\Large JINR preprint E17-96-90}
\end{flushright}

\begin{center}
{\huge Particle Correlations \\ [5mm]
in the Generalized Thomas-Fermi\\ [10mm]
Approach} \\ [5mm]
{\large A.A. Shanenko} \\[3mm]
{\it Bogoliubov Laboratory of Theoretical Physics \\
Joint Institute for Nuclear Research \\
141980, Dubna, Moscow region, Russia \\
e-mail:shanenko@thsun1.jinr.dubna.su}
\end{center}

\vspace{0.3cm}

\begin{abstract}
This paper presents the Thomas-Fermi approach generalized to consider
the particle correlations in many-body systems with non-Coulomb
interaction potentials. The key points of the generalization consist in
using integral formulation and extracting the pair distribution function.
The latter has been found to obey the integral equation which, in the
classical limit, is reduced to the well-known equation of Bogoliubov. So,
the approach presented gives the way of taking into account  interactions in
the quantum many-body systems beyond the weak coupling perturbation theory.
\end{abstract}

\vspace{1cm}

{\large It is known that Thomas-Fermi approach~\cite{zim64} together with its
particular variant for the classical many-body systems, the Debye-Hueckel
method~\cite{har71}, is able to provide satisfactory results while
investigating the correlation phenomena in the media with Coulomb
interparticle potential. The advantage of this approach consists in the
extreme transparency of understanding the physics of the particle
correlations. But up to now the applications of Thomas-Fermi method have
been limited to the situation of the Coulomb type interactions. This paper
presents the generalization of the approach to consider the many-body systems
with constituents interacting by means of an arbitrary relevant potential.

To derive the generalization, let us take the system of some particles
uniformly distributed with density $n$ over a region of volume $V$.
Let the potential of the particle interaction be $\Phi(r)\;.$ Assume
that a point object is placed into this region and interacts
with the  surroundings via the same potential. Further we shall treat the
point of the object location as the origin of the coordinates. We are
interested in the equilibrium structure produced around the origin point.
This structure is specified by quantity $n_{str}( r;\theta,n)$ being
the density of the particles located in the vicinity of point~$\vec r\;,
r\equiv\mid\vec r \mid$.
Here $\theta$ denotes the system temperature. Below we shall use
$n_{str}(r)$ instead of $n_{str}(r;\theta,n)$ to simplify
formulae. According to the first ''item`` of Thomas-Fermi approach,
$n_{str}(r)$ can be calculated with the following condition:
\begin{equation}
\mu_{id} \left( n_{str}( r) \right) + \Phi(r)
       + U(r)=const \;\; (\forall \vec r),
\end{equation}
where $\mu_{id}(\rho)$ is the chemical potential of the ideal gas of the
considered particles at density~$\rho$, and $U(r)$
is the energy of the interaction of the particle being at point
$\vec r$ with the other particles of the system. The quantity in the left
side of~(1) can be named the total particle potential. To calculate
$n_{str}(r)$ with~(1), the connection of $U(r)$ with
$n_{str}(r)$ should be fixed. In the case of the systems with Coulomb
interparticle potential, this is realized by employing the neutralizing
background and Poisson equation~[1,2].  But it is impossible for the case of
non-Coulomb interactions. However, we can follow the other way
based on the obvious integral relation
\begin{equation}
U(r)= \int_{V}\Phi(\mid \vec r-\vec y\mid)\;n_{str}(y) d\vec y\; .
\end{equation}
At point $\vec r_{far}$ which is far enough from the origin
of the coordinates, the total particle potential then takes the
value
\begin{equation}
\mu_{id}(n)+ n\int_{V}\Phi(\mid \vec r_{far}-\vec y\mid) d\vec y
\end{equation}
if, of course, we hold to the reasonable assumption that
$$\lim\limits_{r \to \infty} \Phi(r)=0\,.$$
Note, that we consider $r_{far}\equiv\mid \vec r_{far}\mid \ll R$, where $R$
is the minimal distance between the location point of the object and the
system boundary.
So, any boundary effects can be ignored in our investigation.
Equating~(1) and~(3) we can find
\begin{equation}
\mu_{id}(n) - \mu_{id}\left( n_{str}(r)\right)=
             \Phi(r)+ \int_{V}\bigl(n_{str}(y)-n \bigr)
                 \Phi(\mid \vec r-\vec y\mid)d\vec y\; ,
\end{equation}
where the used relation
\begin{equation}
\lim\limits_{V\to 0}\frac{\int_{V}\left(\Phi(\mid\vec r_{far}-\vec y\mid) -
             \Phi(\mid\vec r-\vec y\mid)\right) d\vec y}
            {\int_{V}\Phi(\mid\vec r-\vec y\mid) d \vec y}\;=\;0
\end{equation}
is fulfilled for a lot of the known potentials, in particular, for the
Coulomb potential as well as for the integrable ones.

Now we should realize the second ''item`` of Thomas-Fermi method. It implies
that using $n_{str}(r)$, we are able to study the space correlations
in the many-body system made of the particles uniformly distributed with
density $n$. The most important quantity calculated in
Thomas-Fermi approach and related with the particle correlations, is usually
thought to be the screening radius. Of course, now we can not exploit this
quantity, for it is tightly connected with Coulomb interparticle potential
and the neutralizing background. But there is one more correlation
characteristic, no less important than the screening radius, that may be
evaluated in the approach discussed. It is the pair distribution function
$g(r;\theta,n)$  for which we have
\begin{equation}
g(r;\theta,n)={n_{str}(r;\theta,n) \over n}.
\end{equation}
Using~(4) and~(6) we readily obtain the following
integral equation for the pair distribution function:
\begin{equation}
\mu_{id}(n) - \mu_{id}\left( n g(r)\right)=
             \Phi(r)+ n\int_{V}\bigl(g(y)-1\bigr)
                 \Phi(\mid \vec r-\vec y\mid) d\vec y\; ,
\end{equation}
where $g(r) \equiv g(r;\theta,n)$. To receive some view to what
extent this equation is adequate, let us explore its classical limit.
In this case we have
\begin{equation}
\mu_{id}(n) - \mu_{id}\left( n g(r)\right)= -\theta\,ln\; g(r).
\end{equation}
Therefore, relation~(7) can be rewritten as
\begin{equation}
 -\theta\,ln\; g(r) =
             \Phi(r)+ n\int_{V}\bigl(g(y)-1\bigr)
                 \Phi(\mid\vec r-\vec y\mid) d\vec y\; .
\end{equation}
Equation~(9) makes it possible to generate the expansion in powers of $n$
for $g(r;\theta,n)$:
$$ g(r;\theta,n)=g_{0}(r;\theta)+n\;g_{1}(r;\theta)+\ldots,$$
where
$$ g_{0}(r;\theta)=exp\left(-\frac{\Phi(r)}{\theta}\right)\; ,$$
$$ g_{1}(r;\theta)=-\frac{1}{\theta} \;
			exp\left(-\frac{\Phi(r)}{\theta}\right) \;
	     \int_{V}\Phi(\mid\vec r-\vec y\mid)\;
         \Bigl(exp\bigl( - \frac{\Phi(y)}{\theta}\bigr)-1\Bigr)\;d\vec y\;.$$
As it is seen, $ g_{0}(r;\theta)$ is in full agreement with the
known result of the calculations in Gibbs canonical ensemble~\cite{bog46}.
But expression for $g_{1}(r;\theta)$ coincides with the true value
\begin{eqnarray}
g_{1}(r;\theta)&=&exp\left(-\frac{\Phi(r)}{\theta}\right)
                 \cdot  \; \nonumber \\
&&\cdot \int_{V}\Bigl(exp\bigl(-\frac{\Phi(\mid\vec r-\vec y\mid)}{\theta}
                                                      \bigr)-1\Bigr)\;
	\Bigl(exp\bigl(-\frac{\Phi(y)}{\theta}\bigr)-1\Bigr)d\vec y
\nonumber
\end{eqnarray}
only at high temperatures. So, equation~(9) gives the valid second virial
coefficient at low temperatures, and the correct second and third ones at
high temperatures. Note, that quite reasonable integral equations for
$g(r)$ derived in the superposition approximation~\cite{hil56}, are
in error beyond the third virial coefficient~\cite{bog46,row68}. Thus, the
use of integral formulation for Thomas-Fermi approach instead of the
differential one and the orientation to the pair distribution function,
enable us to generalize this method to study the case of the non-Coulomb
interactions. But the investigation goal has not been reached yet.

Indeed, equation~(7) gives the possibility to consider particle
correlations in, for example,
nuclear matter, where the nucleon-nucleon potential is of Yukawa type
$exp(-y)/y$, or in the electron gas. But it is not the case for the potentials
which are not integrable due to their behaviour at small separations between
interacting particles.  For instance, if we take potential $1/r^m\;(m>2)$
then the integrals in expressions~(7) and~(9) will not exist. This obstacle
appears because particle correlations have been neglected while
calculating $U(r).$ In particular, in above mentioned reasonings,
quantity $U(r_{far})$ is given with the expression
$$U(r_{far})=n\int_{V}\Phi(\mid \vec r_{far}-\vec y\mid) d\vec y $$
that corresponds to the particle interactions taken in
Hartree approximation. The latter is well-known not to take into account
any correlations. To derive a more correct relation for quantity
$U(r_{far})$, note, that it is nothing else but the interaction energy
of a particle with the other particles uniformly distributed around with
density $n$. Keeping this in mind, we estimate
$U(r_{far})$~\cite{tem68} as
\begin{equation}
U(r_{far})=n\int_{V} g(\mid\vec r_{far}-\vec y\mid)\;
                        \Phi(\mid\vec r_{far}-\vec y\mid)d\vec y \; .
\end{equation}
But now the problem arises how $U(r)$ can be specified, for
we have fixed only its limit value at $\vec r_{far}$. A reasonable way
to do this is to treat quantity
$ g(r)\Phi(r)$ as the effective interaction potential which
should be substituted for $\Phi(r)$ in all the integrals in
expressions~(2)-(5). In this case we have
\begin{equation}
U(r)= \int_{V} g(\mid\vec r-\vec y\mid) \; \Phi(\mid\vec r-\vec y\mid) \;
                                    n_{str}(y) d\vec y\; .
\end{equation}
Further, (1) and~(11) together with~(6) yield the equation
\begin{equation}
\mu_{id}(n) - \mu_{id}\left( n g(r)\right)=
             \Phi(r)+ n\int_{V}\bigl(g(y)-1\bigr) \;
                 g(\mid\vec r-\vec y\mid) \;
                            \Phi(\mid\vec r-\vec y\mid) d\vec y\; .
\end{equation}
It is interesting to compare~(12) with the well-known integral equations
for the pair distribution function of the classical simple liquids.
In such a situation expression~(12) can be rewritten as
\begin{equation}
-\theta\, ln\;g(r)=
             \Phi(r)+ n\int_{V}\bigl(g(y)-1\bigr) \;
           g(\mid \vec r-\vec y \mid) \; \Phi(\mid \vec r-\vec y \mid)
                                         \;  d\vec y\; .
\end{equation}
This is very similar to the equation
\begin{equation}
-\theta\, ln\;g(r)=
             \Phi(r)+ n\int_{V} \;\bigl(g(y)-1\bigr)
             \stackrel{\sim}{\Phi}(\mid \vec r-\vec y\mid) d\vec y\;
\end{equation}
derived in the superposition approximation by Bogoliubov~\cite{bog46}.
Here
$$ \stackrel{\sim}{\Phi}(\mid \vec r-\vec y\mid) \equiv
g(\mid \vec r-\vec y \mid) \Phi(\mid \vec r-\vec y \mid)+
\int_{\mid \vec r-\vec y \mid}^{\infty} \Phi(t)\frac{dg}{dt} dt. $$
Function $\Phi(t)\frac{dg}{dt}$ is oscillating around zero, so its integral,
apparently, does not make an essential contribution into the expression for
$\stackrel{\sim}{\Phi}(\mid \vec r-\vec y\mid)\,. $ So, equations~(13)
and~(14) have to yield the similar qualitative results. According to~(14),
we can expect that $\stackrel{\sim}{\Phi}(r)$ is more accurate estimation
of the  effective interaction potential than $g(r)\Phi(r)$. The replacement
of $g(r)\Phi(r)$ by $\stackrel{\sim}{\Phi}(r)$ in expression~(12), results in
the following integral equation:
\begin{equation}
\mu_{id}(n) - \mu_{id}\left( n g(r)\right) =
             \Phi(r)+ n\int_{V} \;\bigl(g(y)-1\bigr)
             \stackrel{\sim}{\Phi}(\mid \vec r-\vec y\mid) d\vec y\;,
\end{equation}
which gives Bogoliubov equation~(14) in the classical limit. Thus, we have
now derived integral equations~(13) and~(15) that allow us to explore particle
correlations even in the many-body systems with the interaction potentials
behaving at small distances as $1/r^m\;(m>2)\,.$

In conclusion, let us take notice of the most important points of the paper
once more. The article presents the generalization of the Thomas-Fermi method
to investigate particle correlations in the many-body systems with
non-Coulomb interactions of their constituents. The essentials of the
generalization consist in using integral formulation and operating with
the pair distribution function. The paper results in the integral equations
for $g(r)$ which correspond to the different ways of considering particle
interactions. The most accurate of them, (15), is reduced to Bogoliubov
equation~(14) in the classical limit. As to relation~(7), it follows from~(15)
in approximation~$g(r) \approx 1\,.$ At last, integral
equation~(12) can be used as the simplified variant of~(15) being able to
produce the same qualitative picture of particle correlations. Of course,
it is also interesting to investigate other possible evaluations of
$U(r)$ and the corresponding integral equations for the pair distribution
function. This investigation will be presented in the forthcoming paper.

\end{document}